\documentclass[fleqn]{aa}
\usepackage{graphicx, natbib,epsfig,amsmath,amssymb, mathrsfs, txfonts}
\DeclareGraphicsExtensions{.eps, .jpg, .ps}
\usepackage{epstopdf}
\bibpunct{(}{)}{;}{a}{}{,}
\begin{document}

\title{Multiwavelength active optics Shack-Hartmann sensor \\ for seeing and turbulence outer scale monitoring}
\author{P. Martinez\inst{1}}
\institute{Laboratoire Lagrange, UMR7293, Universit\'e de Nice Sophia-Antipolis, CNRS, Observatoire de la C\^ote d\'{}Azur, Bd. de l\'{}Observatoire, 06304 Nice, France} 
\offprints{patrice.martinez@oca.eu}

\abstract
{Real-time seeing and outer scale estimation at the location of the focus of a telescope is fundamental for the adaptive optics systems dimensioning and performance prediction, as well as for the operational aspects of instruments.} 
{This study attempts to take advantage of multiwavelength long exposure images to instantaneously and simultaneously derive the turbulence outer scale and seeing from the full-width at half-maximum (FWHM) of seeing-limited images taken at the focus of a telescope. These atmospheric parameters are commonly measured  in most observatories by different methods located away from the telescope platform, and thus differing from the effective estimates at the focus of a telescope, mainly because of differences in pointing orientation, height above the ground, or local seeing bias (dome contribution).} 
{Long exposure images can either directly be provided by any multiwavelength scientific imager or spectrograph, or alternatively from a modified active optics Shack-Hartmann sensor (AOSH). From measuring simultaneously the AOSH sensor spot point spread function FWHMs at different wavelengths, one can estimate the instantaneous outer scale in addition to seeing.}
{Multiwavelength long exposure images provide access to accurate estimate of $r_0$ and $L_0$ by adequate means as long as precise FWHMs can be obtained. 
Although AOSH sensors are specified to measure not spot sizes but slopes, real-time $r_0$ and $L_0$ measurements from spot FWHMs can be obtained at the critical location where they are needed with major advantages over scientific instrument images: insensitivity to the telescope field stabilization, and being continuously available.}
{Assuming an alternative optical design allowing simultaneous multiwavelength images, AOSH sensor gathers all the advantages for real-time seeing and outer scale monitoring. With the substantial interest in the design of extremely large telescopes, such a system could have a considerable importance.}


\maketitle

\section{Introduction}
The atmospheric seeing is commonly measured by the differential image motion monitor \citep[DIMM,][]{Sarazin90} in most observatories, or by means of alternative seeing monitors, e.g., Generalized Seeing Monitor \citep[GSM,][]{Ziad00}, Multi-Aperture Scintillation Sensor \citep[MASS,][]{Kornilov01}.
Being localized away from the telescope platform, the DIMM delivers seeing estimate that can significantly differ from the \textit{effective} seeing as seen at a telescope focus because of  pointing orientation and/or height above the ground differences, or local seeing bias (dome contribution). The effect of the two latter will substantially expand in the context of the next generation of telescopes: the extremely large telescopes (ELTs).

The evaluation of the seeing is paramount for selecting astronomical sites and/or following their temporal evolution. Likewise, its estimation is fundamental for the dimensioning of adaptive optics (AO) systems and their performance predications. In this context, the outer scale of the turbulence, related as the distance over which the spatial power spectral density of phase distortions deviates from the pure $5/3$ power law at low-frequencies (associated with the Kolmogorov-Obukhov turbulence model), it plays a significant role for AO systems. In particular, the power in the lowest Zernike aberration modes (e.g., tip and tilt) is largely affected, and the knowledge of reliable estimate of $L_0$ is of considerable importance in the future area of the ELTs. As a consequence, both seeing and $L_0$ knowledge virtually drive instrument designs or operational aspects at a telescope, and more emphasis is made to develop accurate real-time seeing and outer scale monitors at the focus of a telescope.

For this purpose various flavors of images can be used at the critical location of the telescope focus: (1/) scientific instrument images, (2/) guide probe images, (3/) active optics Shack-Hartmann images. At the VLT (the Very Large Telescope at the ESO Paranal observatory),  focal planes are equipped with an arm used for acquisition of a natural guide star. The light from this star is then split between a guide probe for an accurate tracking of the sky, and a Shack-Hartmann wavefront sensor used by the active optics to control the shape of the primary mirror.
In this context, seeing estimation from the full-width at  half  maximum of a PSF strongly relies on the exposure time that must be long enough so that the turbulence has been averaged (ensuring that all representations of the wavefront spatial scales have passed through the pupil). This is dependent on telescope diameter and turbulence velocity, though it is commonly admitted that 30 seconds average properly the turbulence, introducing significant FWHM biases otherwise. Since the guide probe has exposure times that are no longer than 50 milli-second, it cannot be used as a seeing/outer scale monitor. 
Active Optics Shack-Hartmann (AOSH) sensor gathers all the advantage for real-time monitoring of these turbulence parameters. While most of the instruments are affected by observational bias (unavailability for a large range of seeing conditions), and are affected by the telescope field stabilization, AOSH delivers continuously real-time images of long exposure spot PSFs (typically 45 seconds) at the same location of scientific instruments.  
AOSH images provide simultaneously various data: slopes, intensities, and spot sizes. When short exposures are used, the information provided by both slopes and intensities (i.e., scintillation) can be used to retrieve $Cn^{2}$ profile using correlations of these data from two separated stars \citep{Robert11, Voyez13}. 
When long exposures are used, from the spot sizes in the sub-apertures, it is feasible to retrieve the atmospheric seeing in the line of sight, with the advantage of being insensitive to the telescope field stabilization \citep{SHsensor}. 
While in most observatories the trend is to compare instrument image quality to DIMM or telescope guide camera FWHM measurements of the seeing in order to estimate $L_0$ \citep[e.g.,][]{Floyd}, in this paper, we propose to use the AOSH sensor system of the telescope as a turbulence monitor to provide accurate seeing estimation directly at the telescope focus, with the additional measurement of the instantaneous turbulence outer scale $L_0$ through multiwavelength exposures. 
For this purpose, an alternative optical configuration of the AOSH system is proposed, and the accuracy of seeing and $L_0$ estimation is analyzed through extensive simulations. 

\section{Analytical treatment}

\subsection{Long-exposure seeing-limited PSF}
The theoretical expression of a long-exposure seeing-limited PSF can be described through the Fourier transform (FT) of its optical transfer function (OTF). The OTF is obtained by multiplying the telescope OTF, denoted $T_{0}(\bf{f})$, by the atmospheric OTF:
\begin{equation}
T_a ({\bf f}) =  \exp [ -0.5 D_{\phi}(\lambda {\bf f})], 
\label{eq:Tf}
\end{equation}
where  ${\bf  f}$ is the angular spatial frequency, $\lambda$ is the  imaging wavelength, and $ D_{\phi}({\bf  r})$ is the phase structure  function \citep{Good85, Roddier81}.  
Equation \ref{eq:Tf} is  appropriate to any turbulence spectrum and any telescope diameter. \\
\noindent The standard theory based on the Kolmogorov-Obukhov model \citep{T61} provides the analytic  expression for the  phase structure function and is expressed by: 
\begin{equation}
D_{\phi}(r) = 6.88 (r/r_0)^{5/3}.
\end{equation}
\noindent This theory describes the shape of the atmospheric long-exposure PSF by a single parameter, the Fried's coherence radius $r_0$ \citep{Fried66}, and the long-exposure OTF is expressed as:
\begin{equation}
T(\textbf{f}) = T_{0} (\textbf{f}) \times exp[-3.44 (\lambda \textbf{f}/r_{0})^{5/3}].
\label{A1} 
\end{equation}
In the case of a large ideal telescope with diameter $D \gg
r_0$ the  diffraction term $T_{0}$ can be neglected. \\ 
However, it is commonly accepted that Eq.~\ref{A1} assumes non-realistic behavior of the low-frequency content of the turbulence model phase spectrum, and it is firmly established that the phase spectrum deviates from the power law at low frequencies \citep{Ziad00, Toko07}. This behavior is described in a first order by an additional parameter, the outer scale $L_0$. 
This additional parameter is introduced by the Von K\`arm\`an  (vK) turbulence  model \citep[e.g.,][]{T61, Ziad00}, and the mathematical definition of $L_0$ is given by:
\begin{equation}
W_{\phi}  ({\bf f}) = 0.0229\;  r_0^{-5/3} \; ( |{\bf f}|^2 + L_0^{-2}  )^{-11/6},
\label{eq:L0}
\end{equation}
\noindent where $W_{\phi} ({\bf f})$ represents the phase distortions. 
In this context, the Kolmogorov-Obukhov model corresponds to $L_0=\infty$, and in the vK model $r_0$ describes the high-frequency asymptotic behavior of the spectrum. 
Physically, $L_0$ is related to the largest size of perturbations, and corresponds to a reduction in the low-frequency content of the phase perturbation spectrum. Even though, the vK is not a verified model, it is established that the phase spectrum does deviate from a power law, and existing experimental data on $L_0$ are interpreted in this sense. $L_0$ does not depend on wavelength and typical values are of the order of 20 m with a scatter that can be as large as few hundred meters \citep{Ziad00, Toko07}. 

In the vK model, the expression for the  phase structure function ($D_{\phi}$) with finite outer scale $L_0$ can be found in \citet[][]{T61} or  \citet[][]{Toko2002}: 
\begin{equation}
\begin{split}
D_{\phi}({\bf  r}) = \frac{\Gamma(11/6)}{2^{11/6} \pi^{8/3}} \left[\frac{24}{5} \Gamma \left(\frac{6}{5} \right) \right]^{5/6} \left( \frac{r_{0}}{L_0} \right)^{-5/3} \\
\times \left[ 2^{-1/6} \Gamma \left( \frac{5}{6} \right) - \left( \frac{2 \pi r}{L_0} \right)^{5/6} K_{5/6} \left( \frac{2 \pi r}{L_0} \right) \right], 
\end{split}
\label{eq:SF}
\end{equation} 
where $K_{5/6}(x)$ is the modified Bessel function of the third kind, and $\Gamma(x)$ is the gamma function.
Putting the vK phase structure function into Eq. \ref{A1}, we obtain the atmospheric long-exposure OTF expression, which now depends on both $\varepsilon_0$ and $L_0$. While it is understood that for finite $L_0$, Eq. \ref{A1} does not go to zero at large spatial frequencies (the FT of $T_a(\textbf{f})$ formally does not exist), but when $r_0 << L_0$ it can be neglected. 

\subsection{Seeing, outer scale, and FWHM}
The Kolmogorov-Obukhov model predicts dependence of the PSF FWHM  $\varepsilon_{0}$  on  wavelength
$\lambda$ and $r_0$:
\begin{equation}
\varepsilon_{0} = 0.976 \; \lambda /r_{0} .
\label{eq1}
\end{equation}
However, as the outer scale of turbulence $L_0$ describes the low-frequency behavior of the turbulence model phase spectrum, it plays a significant role in the image FWHM at the focus of a telescope. 
The image FWHM is  different, and the difference can be as large as 30$\%$ to 40$\%$ in the near-infrared, from the atmospheric seeing that can be measured by dedicated seeing monitors, such as the DIMM \citep[]{Sarazin90}.  \\
The seeing measured by the DIMM is sensitive to small-scale wavefront distortions, and thus provides estimates that are almost independent of the outer scale of the turbulence $L_0$. 
The dependence of atmospheric long exposure resolution on $L_0$ is efficiently predicted by a simple approximate formula (Eq. \ref{toko}) introduced by \citet{Toko2002}, and confirmed by means of extensive simulations \citep{2010A&A...516A..90M, 2010Msngr.141....5M}, where it is emphasized that the effect of  finite $L_0$ is independent of the telescope diameter. The validity of Eq. \ref{toko} has been established in a $L_0$/$r_0$$>$20 and $L_0/D$$\leq$500 domain \citep[][]{2010A&A...516A..90M}, where D is the telescope diameter (the treatment of the diffraction failed for small telescope diameters, D$<$1m). 
As a consequence, long exposure seeing-limited PSF FWHM ($\varepsilon_{vK}$) are directly related to $\varepsilon_0$ (or alternatively $r_0$) and $L_0$ as:
\begin{equation}
\varepsilon_{\rm vK}  \approx \varepsilon_{0} \; \sqrt{1 - 2.183 ( r_{0} / L_{0})^{0.356}}.
\label{toko}
\end{equation}
\begin{figure*}[!ht]
\centering
\includegraphics[width=17.cm]{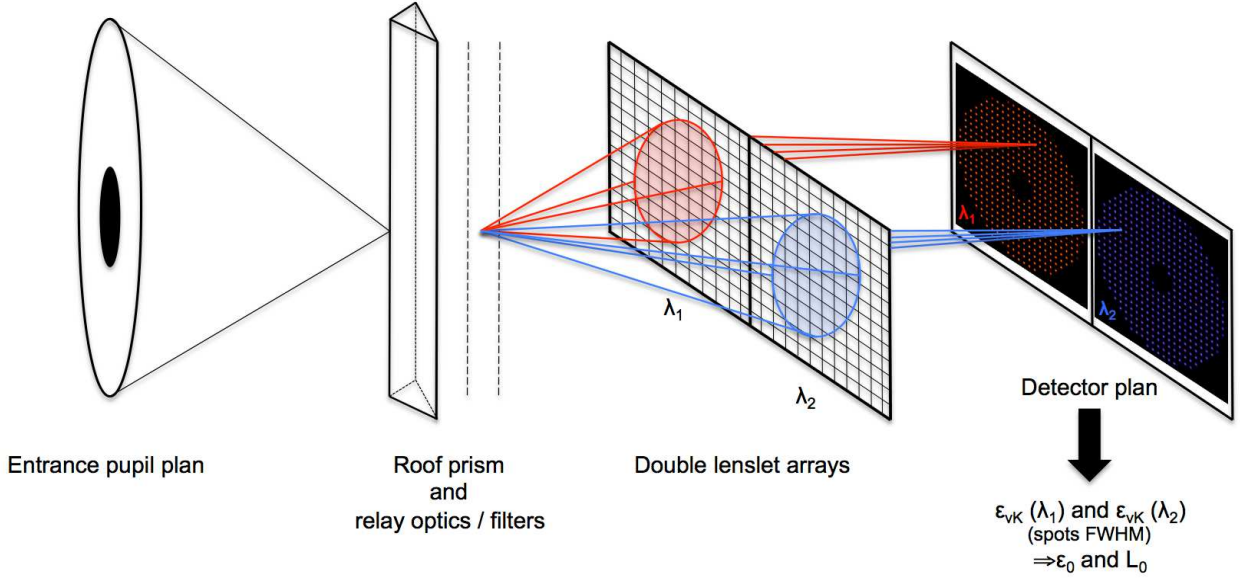}
\caption{Principle of the two-wavelength active optics Shack-Hartmann sensor}
\label{FIG1}
\end{figure*}
Deducing the atmospheric seeing $\varepsilon_{0}$ from the FWHM of a long-exposure seeing-limited PSF at the focus of a telescope requires the correction implied by Eq. \ref{toko} with either the knowledge of $L_0$ (simultaneously measured by any means), or the correction by an a priori $L_0$ value selected from a long term monitoring of the observational site (usually the median value is retained, e.g., $L_0$ = 22 m at Paranal), prior to airmass and wavelength correction. It is trivial to see that these two parameters ($\varepsilon_0$, or alternatively $r_0$, and $L_0$) cannot be simultaneously deduced from a single long exposure seeing-limited PSF FWHM (i.e., a single wavelength exposure provide a unique FWHM estimate, while two parameters are undetermined).

\subsection{Multi-wavelength FWHMs}
A straightforward way for solving the indetermination is to consider a multiwavelength FWHM measurements. In the following, and for the sake of generality $n$ measurements will be assumed, while only two are required to resolve the current problem. In the context of $n$-wavelength long exposure seeing-limited PSF FWHM measurements, Eq. \ref{toko} can be formalized such as:
\begin{equation}
\varepsilon_{\rm vK} ({\bf \lambda_{n}}) \approx \varepsilon_{0} \left( \frac{\lambda_0}{\lambda_n} \right)^{1/5} \times \sqrt{1 - 2.183 \left( \frac{0.976  \times \lambda_0} {\varepsilon_{0} \left( \frac{\lambda_0}{\lambda_n} \right)^{1/5} L_{0}} \right)^{0.356}}, 
\label{tokon}
\end{equation}
\noindent where $\lambda_n$ is the wavelength of the exposure considered, $\lambda_0$ is the wavelength of 500nm, adopted as standard for seeing computation, and $r_0$ has been replaced by $\varepsilon_0$ using Eq. \ref{eq1}. 
Equation \ref{tokon} now provides $n$ different $\varepsilon_{vK}$ estimations, for two unknown parameters ($\varepsilon_0$ and $L_0$), and can be simplified as:
\begin{equation}
\varepsilon_{\rm vK}^{2} ({\bf \lambda_{n}})= A_{n}^{2/5} \times (\varepsilon_{0})^{2} - 2.183 \times A_{n}^{0.328} \times B^{0.356} \times \frac{(\varepsilon_{0})^{1.644}}{(L_{0})^{0.356}},
 \label{tokoF}
\end{equation}
\noindent where $A_n = \lambda_0 / \lambda_n$ and $B = 0.976 \times \lambda_0$. Finally, a simple expression can be obtained:
\begin{equation}
\varepsilon_{\rm vK}^{2} ({\bf \lambda_{n}})= A_{n}^{2/5} \times U - 2.183 \times A_{n}^{0.328} \times B^{0.356} \times V,
 \label{tokoF2}
\end{equation}
\noindent where $U$ and $V$ are equal to $\varepsilon_0^{2}$ and $\varepsilon_0^{1.644} / L_0^{0.356}$ respectively.
A numerical resolution of the system solving for $U$ and $V$ where $n=2$ can be trivially achieved with a standard singular value decomposition method. 

\section{Numerical simulations}
The schematic representation of the modified AOSH sensor considered to provide two-wavelength images is presented in Fig. \ref{FIG1}.
It is based on the standard principle of the AOSH except that at a focal plane downstream of the Shack-Hartmann lenslet array, the light from the telescope pupil is focused at the tip of a roof prism, which splits the light into two parts. These two separated beams propagate through different spectral filters. Finally, the incoming light of these two optical arms is dissected by individual lenslet arrays, which then focus the light onto the detector array. The lenslet array creates a number of separated focal spots of light on the detector.
Belonging to the telescope active optics system, such a system has the advantage of delivering continuously spot images (unaffected by observational bias in contrast to scientific instruments). 
Retrieving atmospheric seeing $\varepsilon_{0}$ (at 500nm) from the FWHM of a unique (single-wavelength) long-exposure seeing-limited PSF image requires the correction implied by Eq. \ref{toko} (prior to airmass and wavelength correction). I recall the reader that it does even concern the case of small size ($d$) AOSH sub-apertures, where $L_{0}$ $\gg$ $d$, and the examination of the $L_0$ influence was also treated \citep{SHsensor}. Accurate seeing estimation by means of a dedicated algorithm and upon adequate calibration is demonstrated in \citet{SHsensor}. 
In the following sub-sections, the details of the simulations involved in the study are explained. 

\subsection{Atmospheric turbulence}
The atmospheric turbulence is simulated with hundreds to thousands uncorrelated phase screens of dimension $4096 \times 4096$ pixels (i.e. 60 meters width) to allow long exposure images ($>30 s$) upon atmospheric conditions. 
The principle of the generation of a phase screen is based on the standard Fourier approach: randomized white noise maps are colored in the Fourier space by the turbulence power spectral density (PSD) function, and the inverse Fourier transform of an outcome correspond to a phase screen realization.
The validity of the atmospheric turbulence statistic has been verified on the simulated phase screens:  the value of the outer scale ($L_{0}$), Fried parameter ($r_{0}$) and seeing of the phase screens have been confirmed by decomposition on the Zernike polynomials and variance measurements over the uncorrelated phase screens. In addition the validity of the long-exposure AOSH image has been verified. In Fig. \ref{model}, we show an example of the simulated AOSH image (based on the ESO VLT AOSH configuration). 

\subsection{Shack-Hartmann model}
\begin{figure}[!ht]
\centering
\includegraphics[width=8cm]{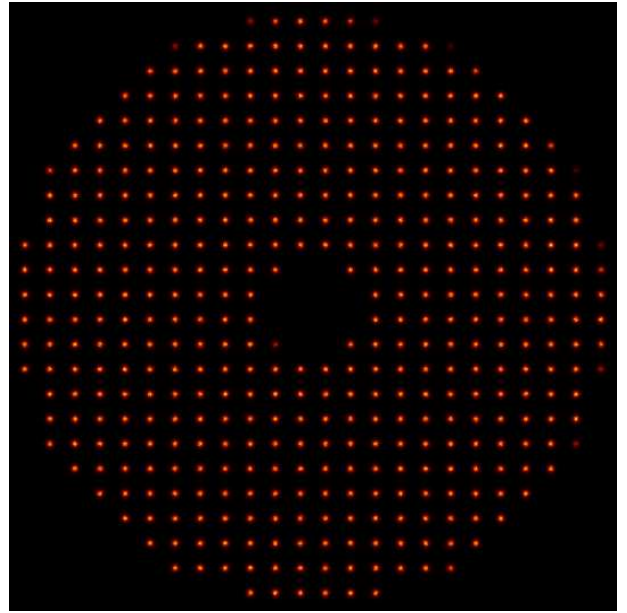}
\caption{Simulated AOSH image based on the VLT AOSH geometry.}
\label{model}
\end{figure}

Simulations are based on a diffractive Shack-Hartmann model that reproduces the VLT AOSH geometry, mimicking the 24 sub-apertures across the pupil diameter of the VLT active optics, with 22 pixels per sub-aperture, and 0.305$\arcsec$ pixel scale ($d = D/24 = 0.338$ m). The validity of the AOSH model has been verified through several aspects such as the plate scale, spot sizes, slope measurements, and phase reconstruction. The AOSH paradigm is presented in Fig. \ref{model}. 
Background estimation is performed on a corner of the image without spots, and hot pixels are set to the background.
The cleanest and un-vignetted spots are selected in each frame for the analysis. These extracted spots are oversampled by a factor two, re-centered and averaged. The averaging reduces the influence of potential local CCD defects, though these are not included in the simulation. In practice, the averaged spot is based on hundreds of selected spots (more than 400 spots over 526 available are usually selected for the analysis). 
The multiwavelength AOSH simulator provides 4 images at 4 different wavelengths (0.5, 0.55, 0.6, and 0.65 $\mu$m), while only two are used and analyzed (0.5 and 0.55 $\mu$m). The choice of the wavelength values used in simulation is arbitrary and its optimization is beyond the scope of the present study. While an optimal combination can certainly be found by considering system and telescope operational aspects (e.g., photon noise, bandpasses, sub-system characteristics and pertaining constraints, etc.), once the wavelengths are known and corresponding FWHMs correctly estimated, seeing and outer scale are accurately retrieved regardless the wavelength combination selected.

\subsection{Extraction of the FWHM}
The algorithm used to extract the FWHM from the AOSH sensor spots has been proposed by \citet{Toko07} and extensively analyzed in \citet{SHsensor}. It is based on the long-exposure spot PSF profile defined in Eq. \ref{A1}.  Its principle is the following: the modulus of the long-exposure optical transfer function of the averaged spot is calculated and normalized. It is then divided by the square sub-aperture diffraction-limited transfer function $T_0(\bf{f}$): 
\begin{equation}
T_{0}(\bf{f}) = (1-[\lambda \bf{f_x}/d]) \times (1 - [\lambda \bf{f_y}/d]).
\label{diff1}
\end{equation}
In the case of a large ideal telescope with diameter $D \gg r_0$ the diffraction term $T_{0}$ of Eq. \ref{A1} can be neglected, while in the case of AOSH sub-apertures of size $d$ it cannot ($d\approx r_0$). 
At this stage the cut of the $T(\bf{f})$ along each axis can be extracted and fitted to the exponential part of Eq. \ref{A1} to derive a single parameter $r_0$, or equivalently Fourier transformed to derive the FWHM of the resulting spot PSF profile using a 2-dimensional elliptical Gaussian fit (a Moffat or a 10th order polynomial fits can be alternatively selected). The orientation of the long and small axis of $T(\bf{f})$ is found by fitting it with a 2-dimensional elliptical Gaussian. 

\subsection{Solving $\varepsilon_0$ and $L_0$ from FHWMs}
Various methods for solving a set of $n$ linear equations in $n$ unknowns are available, e.g., least squares fitting, LU (lower/upper) decomposition, etc. 
In practice, the standard LU decomposition has been successfully used to solve our square system of linear equations. 
The programming language is IDL using the linear algebra library package \textit{LAPACK}. 
\begin{figure*}[!ht]
\centering
\includegraphics[width=8.9cm]{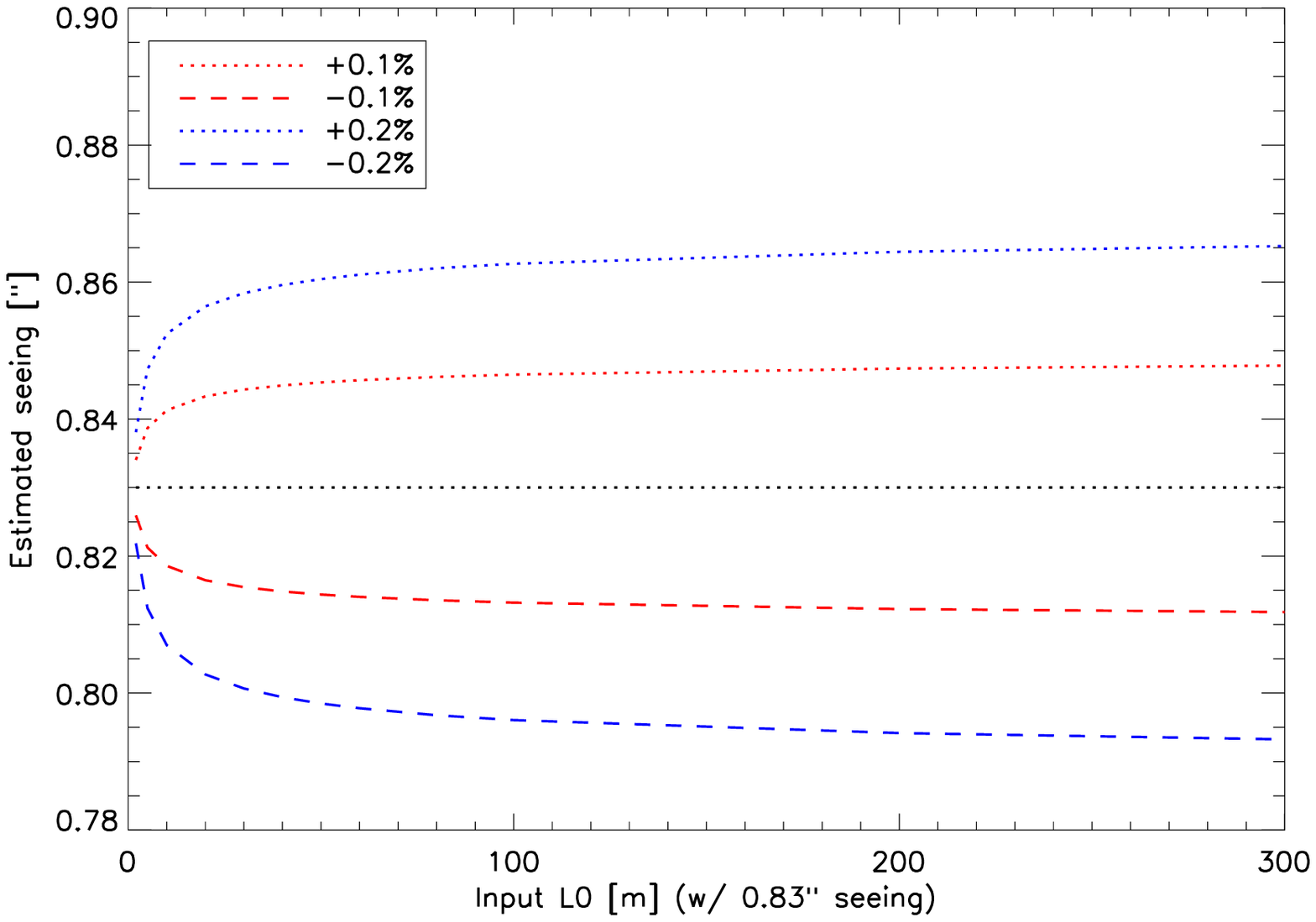}
\includegraphics[width=8.9cm]{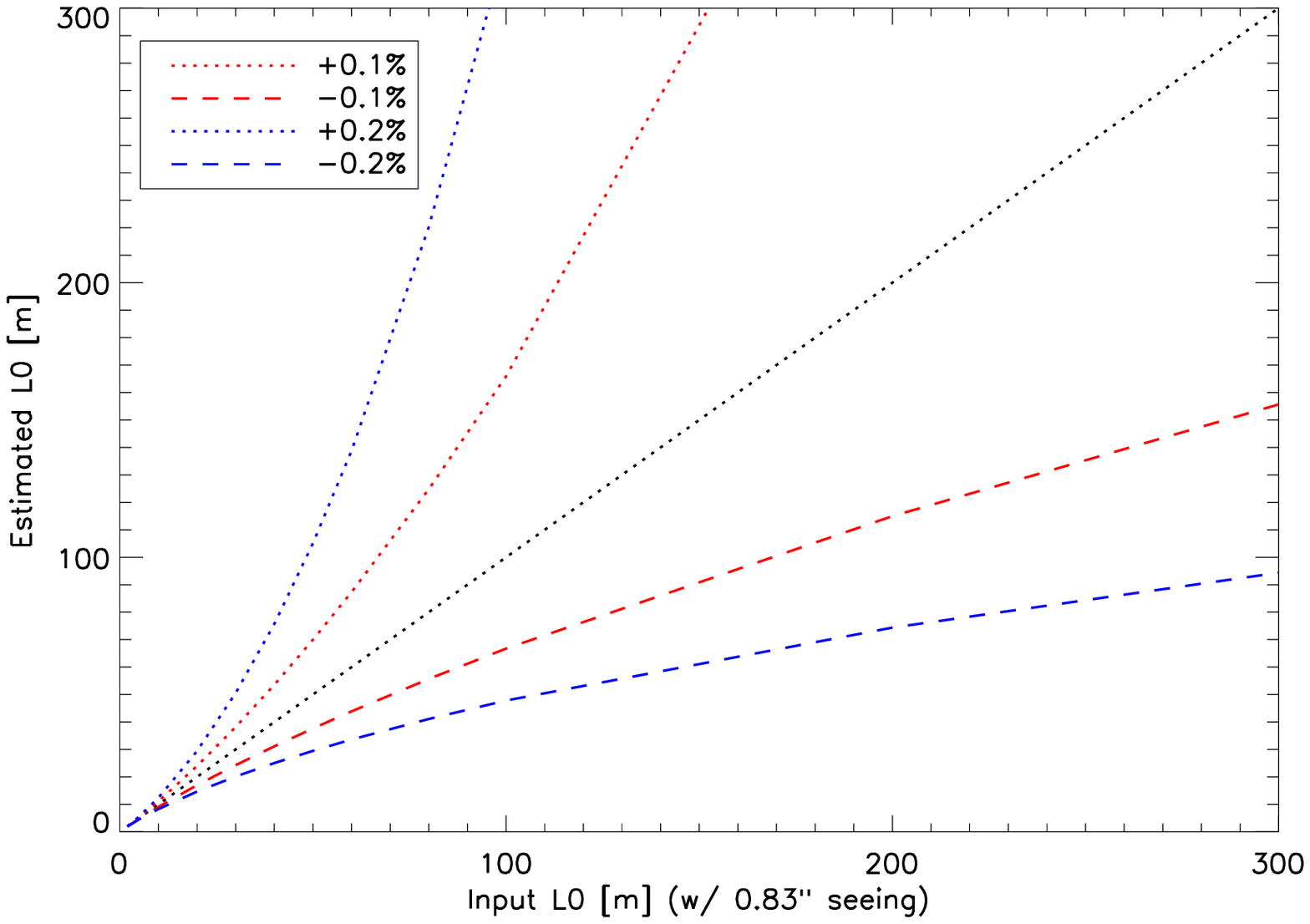}
\caption{Impact of a differential error in the estimation of the FWHM measured at 0.5 and 0.55 $\mu$m with the two-wavelength AOSH sensor on the seeing estimate (left) and turbulence outer scale estimate (right). The dashed and black lines represent the case where no error is applied. } 
\label{FIG3}
\end{figure*}
\section{Results and discussion}
\subsection{General results}
\label{results}
The series of test conducted to determine the ability of a multiwavelength AOSH sensor to sort out seeing and turbulence outer scale values are twofolds: (1/) under a specific simulated seeing condition (0.83 $\arcsec$, ESO Paranal observatory median seeing value as measured by the DIMM at 6 meter height from the ground), several turbulence outer scale conditions are simulated, and simulated/estimated parameters can be compared; (2/) for a particular generated turbulence outer scale value, several seeing conditions (ranging from 0.6 to 1.8 $\arcsec$) are simulated. Estimated and simulated parameters are then again compared. This latest test has been repeated for two outer scale values (22 and 30 meter, where 22 m is the Paranal observatory median value).  \\
At this stage, two aspects should be pointed out, both being related to sampling issue. 
(1/) Larger outer scale values than 30 meters cannot in principle be considered as the simulated atmospheric phase screens are physically limited to 60 meters width. It was indeed observed that above roughly 40 meters phase screen sides, the outer scale estimation failed to be accurate. 
(2/) There is a catch in matching the simulated AOSH sensor to the one of the VLT: the pixel scale is crude (0.305$\arcsec$). This has been pointed out in \citet{SHsensor}: seeing estimates under 0.6 $\arcsec$ failed to be accurate, while probably seeing values above 0.6 $\arcsec$ are still affected to a certain extent by under sampling effect. 
This point will be further discussed in the Section \ref{pixel}, were the AOSH geometry is modified (by simulating AOSH sensor with various sub-apertures configurations but identical telescope pupil footprint in pixel, in order to increase or decrease the pixel sampling, e.g.,  6$\times$6, 9$\times$9, 18$\times$18, 24$\times$24, and 48$\times$48 sub-apertures). \\
Table \ref{Table1} summarizes the results obtained under various turbulence outer scale values coupled with 0.83$\arcsec$ seeing conditions. Both seeing and outer scale values are well estimated from the AOSH images. The seeing is accurate at a $10^{-2}$ level, while the outer scale is fairly estimated with a maximum error margin of 2 meters. 
Table \ref{Table2} presents the results under various seeing conditions with two outer scale values (22 and 30 meters). For the 22 m outer scale set of data, the mean value of the estimated outer scale is 24.5 m with a standard deviation of 7.9 m, while for the 30 m outer scale set of data, the mean value is 28.7 m with a standard deviation of 5.7 m. 
In all cases, the seeing estimate is fairly good, accurate at a few $10^{-2}$ level. In addition, from Table \ref{Table2} it is observable that the better the seeing, the worse the outer scale estimates. Increasing the seeing estimate accuracy is therefore fundamental for increasing the accuracy of the outer scale estimation. This is likely due to the pixel scale issue, that directly impacts the accuracy of the FWHM estimates. This point is addressed in the next sub-section.

\begin{center}
\begin{table}[t]
\centering
\begin{tabular}{c|c|c|c}
\hline
\multicolumn{2}{c|}{Simulated parameters} & \multicolumn{2}{c}{Measured parameters} \\
\hline
$L_0$ [m] & Seeing [$\arcsec$] & $L_0$ [m] & Seeing [$\arcsec$] \\
\hline
20.0 & 0.830 & 21.75 & 0.829 \\
30.0 & 0.830 & 28.02 & 0.833 \\
40.0 & 0.830 & 39.96 & 0.823\\
\hline
\end{tabular}
\caption{Seeing and $L_0$ estimations for various $L_0$ values under 0.83 $\arcsec$ seeing conditions.}
\label{Table1}
\end{table}
\end{center}
\begin{center}
\begin{table}[t!]
\centering
\begin{tabular}{c|c|c||c|c}
\hline 
& \multicolumn{4}{c}{Input $L_0$}  \\
\cline{2-5}
& \multicolumn{2}{c||}{22 m} & \multicolumn{2}{c}{30 m} \\
\cline{2-5}
& \multicolumn{2}{c||}{Measured parameters} &  \multicolumn{2}{c}{Measured parameters}\\
\hline
Input seeing [$\arcsec$] & $L_0$ [m] & Seeing [$\arcsec$] &  $L_0$ [m] & Seeing [$\arcsec$]  \\
\hline
0.600 &       42.29  &  0.663   & 19.90 & 0.617 \\
0.700 &       17.85  &   0.715    &   32.15 & 0.691 \\
0.800 &      35.78  &       0.769   &  26.83 & 0.832 \\
0.900 &      14.37   &        0.936 &   20.83 & 0.919 \\
1.000 &       29.17   &      0.979  &      36.16 &  0.983  \\
1.100 &      25.85    &      1.089   &   37.42 & 1.082 \\ 
1.200 &      29.05    &      1.177  &    32.22 & 1.192 \\ 
1.300 &     22.61     &     1.298  &   34.61 & 1.286 \\   
1.400 &    18.96      &   1.415  &     30.33 & 1.396 \\   
1.500 &     22.13     &   1.500  &       26.36& 1.509  \\ 
1.600 &   18.91       &    1.615  &   25.87& 1.611  \\     
1.700 &   17.72       &    1.724 &    22.46 & 1.724 \\    
1.800 & 23.95 &   1.793     &   27.39 & 1.802 \\  
\hline
\end{tabular}
\caption{Seeing and $L_0$ estimations for various seeing conditions.}
\label{Table2}
\end{table}
\end{center}
\begin{center}
\begin{table*}[t]
\centering
\begin{tabular}{c|c|c|c}
\hline 
&  & \multicolumn{2}{c}{Measured FWHM [$\arcsec$] }  \\
\hline
 AOSH geometry [sub-apertures] & Pixel sampling [$\arcsec$] & 0.5 $\mu$m & 0.55 $\mu$m \\
\hline
48$\times$48 & 0.610 & 0.700 & 0.723 \\
24$\times$24 &0.305 & 0.615 & 0.635 \\
18$\times$18 & 0.228 & 0.573 & 0.598 \\
9$\times$9 & 0.114 & 0.520 & 0.509 \\
6$\times$6 & 0.076  & 0.508 & 0.494 \\
\hline
\end{tabular}
\caption{Seeing is 0.6 $\arcsec$ and $L_0$ is 22 meters. The theoretical FWHM is equal to 0.470 $\arcsec$ and 0.455 $\arcsec$ at 0.5 $\mu$m and 0.55 $\mu$m respectively.}
\label{Table3}
\end{table*}
\end{center}

\vspace{-25mm}
\subsection{Sensitivity analysis}
Precise FWHM estimation from the spot images is mandatory to deliver accurate seeing and outer scale monitoring with the AOSH sensor.
This is of a major importance for the outer scale estimate more than for the seeing itself. This is observable in Fig. \ref{FIG3} where the impact of a differential error in the calculation of the two FWHMs is appraised. The evaluation of the FWHMs from spot images are not perfect, and some errors are committed during the calculation for various reasons: internal system aberrations, spot under or poor sampling, insufficient signal-to-noise, etc. However, it is likely that this unavoidable amount of error involving in such estimation will apply almost similarly for both wavelength measurements, and the impact on the seeing and outer scale estimates will therefore be small as these errors committed at the two wavelengths will roughly compensate for each other. Hence, only the differential error between the two wavelengths matters. This is precisely what is addressed in the following. 
Using Eq. \ref{toko} and Eq. \ref{tokoF2} it is straightforward to compute the seeing and outer scale values knowing the theoretically corresponding FWHMs at 0.5 and 0.55 $\mu$m. The test was done for a 0.83 $\arcsec$ seeing and various outer scale values by introducing a small error in the theoretical FWHM at 0.5$\mu$m (the FWHM computed at 0.55$\mu$m is left unaffected). The impact of this error on the resolution of the multiwavelength system (Eq. \ref{tokoF2}) is then plotted in Fig. \ref{FIG3}, where the impact on the seeing estimate (left) and outer scale estimate (right) are shown. \\
The impact is not significant for the seeing evaluation, but the outer scale. The seeing estimation is mainly affected for low outer scale values ($< 30 m$) and the evolution almost stabilizes for large outer scale values ($>100 m$). The impact on the outer scale estimate has an opposite behavior. Small outer scale values are less sensitive to FWHM error than large values. The departure from the expected value (in dashed black line, i.e., with no error) can be significant, and degrades further with the outer scale. While a reasonable differential error in the estimation of the two-wavelength FWHMs has little impact on the seeing estimation, the impact on the outer scale can be considerable. This raise the importance of either developing a dedicated and accurate algorithms to precisely evaluate FWHMs from AOSH sensor spots upon adequate calibration \citep{SHsensor}, or developing an optimized optical design for the multiwavelength AOSH sensor struggled for high sampling of the spots on the detector. 

\subsection{Sub-aperture sampling}
\label{pixel}
The sampling effect is analyzed by modifying the AOSH pattern. While the footprint in pixel of the lenslet array is left unmodified, the number of sub-apertures is varying to allow changing the number of pixel per sub-aperture. Various configurations are tested (from 6$\times$6 to 48$\times$48 sub-apertures), where the FWHMs can be compared to predictions by Eq. \ref{toko}. The results presented in Table \ref{Table3} are unambiguous: the higher the sampling, the more accurate the FWHM. Sub-aperture sampling can be calibrated but to a certain extent, and the effect is not linear with the seeing. Poor sub-aperture sampling calibration will not be as efficient as for a mild situation. This is an key aspect in the AOSH parameter space to consider when designing a multiwavelength AOSH system. 

\section{Conclusion}
Although AOSH sensors are specified to measure not spot sizes but slopes, the principle of the multiwavelength active optics Shack-Hartmann sensor is demonstrated and emerges as a potential efficient system for real-time seeing and instantaneous turbulence outer scale monitoring. It offers direct access to important turbulence parameters at the precise location where we need them (at the focus of the telescope). Multiwavelength AOSH delivers long-exposure PSFs without being affected by any observational bias in contrast to scientific instruments, and being insensitive to the telescope field stabilization. 
\newpage
In addition, the as proposed AOSH design has the advantage of permitting the calibration of the internal aberrations present in the system/telescope downstream of the roof prism, it precludes their contribution to the spot FWHMs.
The present study shows that accurate seeing and turbulence outer scale estimates can reasonably be achieved with nowadays systems, while the reliability of such a system can trivially be increased with a better sampling of the spot PSFs. Assuming an alternative optical design allowing simultaneous multiwavelength images, the active optics Shack-Hartmann sensor is a good candidate for seeing and outer scale monitoring, especially in the vicinity of the ELT area.


\end{document}